\documentclass[aps,10pt,twocolumn,prd,showpacs,showkeys,preprintnumbers,superscriptaddress,nobibnotes,floatfix,longbibliography,notitlepage,nofootinbib]{revtex4-1}
\pdfoutput=1

\usepackage{amsmath}
\usepackage{amsfonts}
\usepackage{amssymb}
\usepackage{mathrsfs}
\usepackage{cancel}
\usepackage{accents}
\usepackage{mciteplus,slashed}
\usepackage{amssymb,cancel,amsmath,relsize}
\usepackage{mathrsfs} 
\usepackage{dcolumn}
\usepackage{bm}
\usepackage[caption=false]{subfig} 
\usepackage{appendix}
\usepackage{physics}
\usepackage{feynmp-auto}
\usepackage[T1]{fontenc}	
\usepackage{csvsimple}
\usepackage{hyperref}
\usepackage[capitalise]{cleveref}
\usepackage{booktabs}
\usepackage{graphicx}
\usepackage{mathrsfs}
\usepackage[utf8]{inputenc}
\usepackage[dvipsnames]{xcolor}
\hypersetup{
    colorlinks,
    linkcolor={red!50!black},
    citecolor={blue!50!black},
    urlcolor={blue!80!black}
}
\usepackage[normalem]{ulem}
\usepackage{cleveref}
\usepackage{fancybox}

\newcommand{\Ztwo}{\ensuremath{\mathcal{Z}_2}}
\newcommand{\ZDM}{\ensuremath{\mathcal{Z}_2^{\rm (DM)}}}
\newcommand{\ZDW}{\ensuremath{\mathcal{Z}_2^{\rm (DW)}}}

\graphicspath{{Figures/}}

\begin{document}

\title{Quantum Gravity Effects on Fermionic Dark Matter and Gravitational Waves}

\author{Stephen F. King}
\affiliation{School of Physics and Astronomy, University of Southampton, Southampton SO17 1BJ, United Kingdom}

\author{Rishav Roshan}
\affiliation{School of Physics and Astronomy, University of Southampton, Southampton SO17 1BJ, United Kingdom}

\author{Xin Wang}
\affiliation{School of Physics and Astronomy, University of Southampton, Southampton SO17 1BJ, United Kingdom}

\author{Graham White}
\affiliation{School of Physics and Astronomy, University of Southampton, Southampton SO17 1BJ, United Kingdom}

\author{Masahito Yamazaki}
\affiliation{Kavli IPMU (WPI), UTIAS, The University of Tokyo, Kashiwa, Chiba 277-8583, Japan}
\affiliation{Center for Data-Driven Discovery, Kavli IPMU (WPI), UTIAS, The University of Tokyo, Kashiwa, Chiba 277-8583, Japan}
\affiliation{Trans-Scale Quantum Science Institute, The University of Tokyo, Tokyo 113-0033, Japan}

\begin{abstract}

We explore the phenomenological consequences of breaking discrete global symmetries in quantum gravity (QG). We extend a previous scenario where discrete global symmetries are responsible for scalar dark matter (DM) and domain walls (DWs), to the case of fermionic DM, considered as a feebly interacting massive particle, which achieves the correct DM relic density via the freeze-in mechanism. Due to the mixing between DM and the standard model neutrinos, various indirect DM detection methods can be employed to constrain the QG scale, the scale of freeze-in, and the reheating temperature simultaneously. Since such QG symmetry breaking leads to DW annihilation, this may generate the characteristic gravitational wave background, and hence explain the recent observations of the gravitational wave spectrum by pulsar timing arrays. This work therefore highlights a tantalizing possibility of probing the effective scale of QG from observations.

\end{abstract}

\maketitle
\tableofcontents

\section{Introduction}

Discrete global symmetries often play a role in many theories beyond the Standard Model (SM), such as the dark matter (DM) and neutrino mass models. While it is theoretically tempting to invoke global symmetries,
recent developments in the swampland program \cite{Vafa:2005ui,Ooguri:2006in}
suggest that exact global symmetries should be broken \cite{Banks:1988yz,Banks:2010zn,Harlow:2018tng} in theories of quantum gravity (QG). 
Since it is very challenging to find any experimental test of QG, it is an important problem to explore the phenomenological consequences of global-symmetry breaking originating from QG.
In the literature, the dimensional scale associated with the operators that break the symmetry is often associated with the Planck scale $M^{}_{\rm Pl}$~\cite{Addazi:2021xuf}. 
However, the size of the symmetry-breaking effects may be suppressed by non-perturbative effects, leading to an effective breaking scale many orders of magnitude higher than $M^{}_{\rm Pl}$.

In a recent Letter~\cite{King:2023ayw}, we 
proposed that QG effects can be responsible for the instabilities of both DM and domain walls (DWs) through the explicit breaking of discrete symmetries. We especially considered a simple model with two singlet scalar fields and two $\Ztwo$ symmetries, one being responsible for DM stability, and the other spontaneously broken and responsible for DWs; both $\Ztwo$ symmetries are assumed to be explicitly broken by QG effects by operators at the same mass dimension and with the same effective Planck scale. We showed that this hypothesis led to observable gravitational wave (GW) signatures from the annihilation of DWs, which are correlated with the decaying DM signatures constrained by cosmic microwave background (CMB) observations. We also analyzed how the recent GW spectrum observed by several pulsar timing array projects can help constrain these effects, possibly providing an explanation for the results of the North American Nanohertz Observatory for Gravitational Waves (NANOGrav)~\cite{NANOGrav:2023gor, NANOGrav:2023hvm}, the European PTA~\cite{Antoniadis:2023ott, Antoniadis:2023zhi}, the Parkes PTA~\cite{Reardon:2023gzh} and the Chinese PTA~\cite{Xu:2023wog}.

\begin{figure}[t!]
    \centering
    \includegraphics[width=0.9\linewidth]{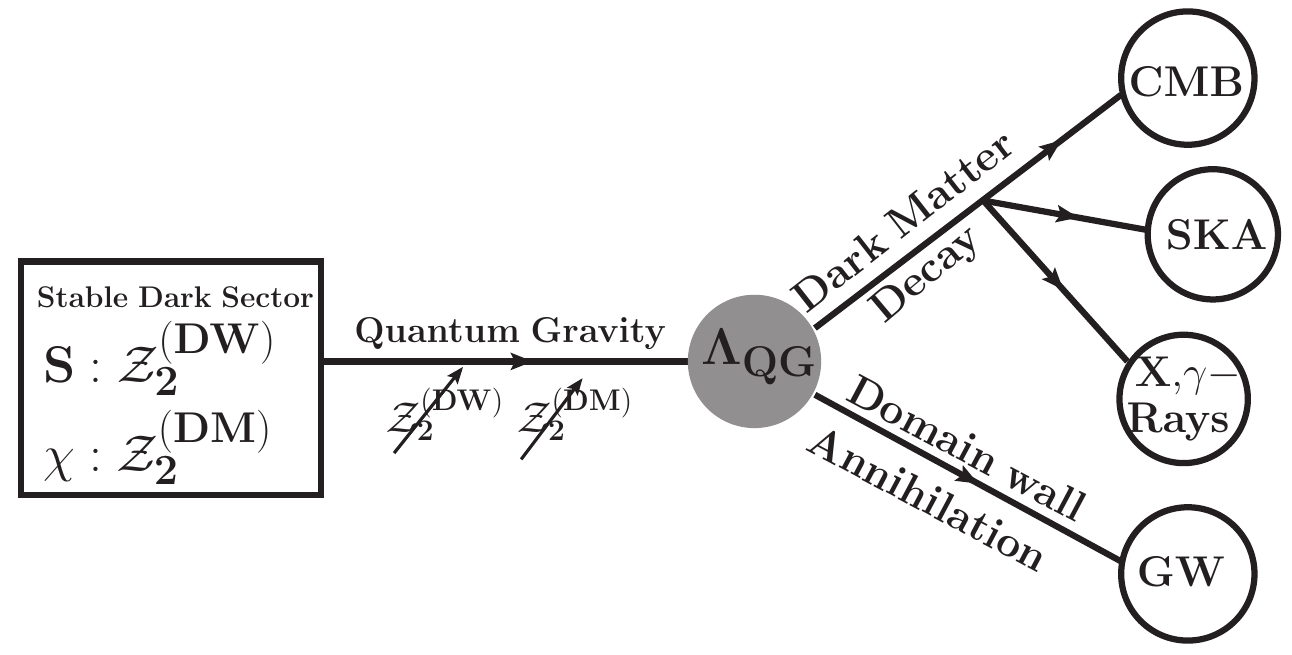}  
    \caption{Schematic of how the DM indirect detections and GW observatories can provide independent witnesses of the common QG scale.}
    \label{fig:schematic}
\end{figure}

In this paper, we extend the results in~\cite{King:2023ayw} concerning the scalar DM to the case of fermionic DM, which is odd under a $\mathcal{Z}_2^{\text{DM}}$ symmetry. Meanwhile, we also introduce a scalar field that is odd under a $\mathcal{Z}_2^{\text{DW}}$ symmetry.  The scalar field acquires a vacuum expectation value (VEV) $v^{}_s$ and breaks the $\Ztwo^{\text{DW}}$ symmetry in the early Universe, generating DWs in the process. Similar to our previous work~\cite{King:2023ayw}, both $\Ztwo$ symmetries are approximate global symmetries inevitably broken by QG effects, leading to the decays of DM and the observable GW signatures from the annihilation of DWs. Then one could expect the DM indirect detections and GW observatories can provide independent witnesses of the common QG scale. Our framework is depicted by the schematic shown in \cref{fig:schematic}. 

It is worth emphasizing that we have several motivations to consider the fermionic DM extension, which are as follows. First, fermionic DM is also well-motivated in string theories, e.g.\ in F-theory GUT models. Second, different from the scalar DM which is assumed to be the weakly interacting massive particle (WIMP)~\cite{Kolb:1990vq, Jungman:1995df, Bertone:2004pz, Feng:2010gw, Arcadi:2017kky, Roszkowski:2017nbc} in our previous work, there are good reasons to consider the fermion as other DM candidates, e.g. the feebly interacting massive particle (FIMP)~\cite{Hall:2009bx}, which brings us into the {\it freeze-in} scenario, as we will discuss below. It should also be mentioned that in string theories one of the motivations for freeze-in actually comes from moduli and modulinos, see e.g.\ ref.~\cite{Hall:2009bx}. Third, the fermionic DM considered in the present paper can have Yukawa interactions with the SM neutrinos, leading to observable effects on CMB, diffuse X/$\gamma$-ray background, radio signals, and neutrinoless double beta decays, which can help constrain the QG scale.

The presence of DM, a mysterious, non-luminous, and non-baryonic form of matter, is strongly suggested by experiments through the precise measurement of  CMB anisotropies \cite{Aghanim:2018eyx}. While the cosmological shreds of evidence are solely based on the gravitational interactions of DM, its particle nature is still unknown. As is well known, the SM of particle physics fails to provide an authentic DM candidate, which motivates us to look for the physics beyond the Standard Model (BSM). 

Several BSM proposals are available in the literature that try to explain the particle aspect of DM. The most popular one is the WIMP paradigm~\cite{Kolb:1990vq, Jungman:1995df, Bertone:2004pz, Feng:2010gw, Arcadi:2017kky, Roszkowski:2017nbc}, where the DM is considered to be thermally produced from the SM bath in the early Universe. It is interesting to point out that if such particles have masses and couplings around the electroweak (EW) scale, they give rise to correct DM abundance after their thermal freeze-out. The same interaction also results in the DM-nucleon scattering with the possibility of detecting them in direct detection experiments like XENON 1T~\cite{XENON:2018voc}, LUX~\cite{LUX:2016ggv}, PandaX-II~\cite{PandaX-II:2017hlx}, etc.
While the null detection of WIMP-type DM in these experiments does not necessarily rule it out, we are
definitely better motivated to consider other DM paradigms where the interaction strength of DM can be much lower than the scale of weak interaction. One such possibility is to consider the DM as a FIMP~\cite{Hall:2009bx}. In such a scenario, the DM never enters into thermal equilibrium with the SM bath. While the DM has negligible initial abundance in the early Universe, it can be produced from out-of-equilibrium decays or annihilation of the bath particles. Here, the DM abundance slowly \emph{freezes-in} from a negligible initial abundance to the observed DM abundance.

If there exist renormalizable interactions between the FIMP and the SM bath, the production of DM is effective at the lowest possible temperature. If the decaying particle is in thermal equilibrium with the bath then the dominant DM production occurs when the temperature of the bath equals the mass of the decaying particle. These types of freeze-in scenarios are known as infra-red (IR) freeze-in~\cite{Bernal:2017kxu, Biswas:2016bfo, Datta:2021elq,  Bhattacharya:2021jli}.  Nevertheless, if the DM and the SM sectors are coupled only through higher dimensional operators (of dimension $d > 4$), the DM production is effective at high temperatures and very sensitive to initial history like the reheating temperature of the Universe. Due to the higher dimensional nature of such interactions, DM production only happens via scattering, especially at a temperature above the EW scale. This particular scenario is known as the ultra-violet (UV) freeze-in~\cite{Elahi:2014fsa,Biswas:2019iqm,Barman:2020plp,Barman:2021tgt}. 

It may also happen that a realistic FIMP scenario is the mixture of both IR and UV freeze-in where after the phase transition like the one occurring at the EW scale, the DM can have renormalizable interactions with the SM bath. However, if DM mass is much higher than the scale of such phase transitions, its production will be dominated by UV freeze-in only. Moreover, as a result of their feeble interactions with the visible sector, testing or experimentally observing a FIMP DM, particularly at colliders or fixed-target experiments, is exceedingly difficult. However, they can still be tested via indirect searches. Due to their feeble interactions, they are expected to have a very long lifetime if allowed to decay. In general, X-ray and $\gamma$-ray observatories act as powerful probes to light freeze-in DM. On the other hand, if the DM decays during or after the recombination, it can leave an imprint on the CMB power spectrum. 
Finally, the decay of DM could produce $e^+e^-$ pairs, which undergo energy loss via electromagnetic interactions and leave radio-wave signals when traveling through the interstellar medium.

The rest of this paper is organized as follows. In \cref{sec:QG}
we discuss the swampland global symmetry conjecture, its refinement, and its phenomenological consequences. In \cref{sec:simplified} we introduce our simplified models involving 
fermionic DM and another scalar field creating DWs. In \cref{sec:DM} 
we discuss the phenomenology of the freeze-in DM.
In \cref{sec:indirect} we discuss the possibility of determining the QG scale 
by indirect detection experiments. 
In \cref{sec:DW} we turn to GWs generated by annihilation of DWs.
In \cref{sec:combined} we synthesize our analysis of DM and GWs.
We end this paper in \cref{sec:conclusion} with concluding remarks.

\section{Swampland Global Symmetry Conjectures} \label{sec:QG}
For decades effective field theories (EFTs) have been a theoretical framework
that has guided particle-physics model builders looking for signatures of new physics.
In this approach, one simply writes down all the possible terms consistent with the 
(gauge and global) symmetries of the problem, and the higher-dimensional operators represent potential signatures for new physics, the size of which can be constrained by experiments.

There has been increasing evidence, however, that there are limitations of this approach:
the situation is different once we include gravity and require that the EFT in question have a UV completion in a suitable theory of QG, such as string theory. While we still are very far from understanding the full power of the UV physics in constraining a given EFT, we can try to formulate a set of necessary (but not necessarily sufficient) conditions for the EFT to have a UV completion. Such conditions are called \textit{swampland conjectures}, where a swampland refers to a set of EFTs that cannot be embedded in theories of QG \cite{Vafa:2005ui,Ooguri:2006in}. There have recently been tremendous developments in the swampland program, see e.g.\ \cite{Brennan:2017rbf,Yamazaki:2019ahj,Palti:2019pca,vanBeest:2021lhn} for summaries.

One of the most famous (and arguably most well-established) swampland conjectures
is the \emph{Swampland Global Symmetry Conjecture (SGSC)} \cite{Banks:1988yz,Banks:2010zn,Harlow:2018tng}:

\medskip

\noindent
{\bf Swampland Global Symmetry Conjecture}:
There exists no exact global symmetry in EFTs arising from UV theories of QG. 
This statement applies to all global symmetries, Abelian or non-Abelian, continuous or discrete.

\medskip

For decades this conjecture has been argued from multiple different angles~\cite{Banks:1988yz,Banks:2010zn,Harlow:2018tng},
and is often considered as one of the most established swampland conjectures in the literature.\footnote{
As already noted in our previous paper \cite{King:2023ayw}, 
the conjecture assumes weakly-coupled Einsteinian gravity, and hence do not necessarily apply to more non-standard theories of gravity. Relatedly, the no-global symmetry conjecture can actually be violated in QG theories involving ensemble averages, see, e.g.\ refs.~\cite{Antinucci:2023uzq,Ashwinkumar:2023jtz} for recent discussions.}
SGSC also follows as a corollary of another famous swampland conjecture,
the weak gravity conjecture \cite{Arkani-Hamed:2006emk}, which in turn is closely related to another swampland conjecture,
the distance conjecture \cite{Ooguri:2006in}.  We will hence assume this conjecture in the rest of this paper.

Given the fundamental importance of the SGSC, it is natural to ask if we can extract any useful consequences in EFTs, in particular in the search for BSM physics. 
We quickly point out, however, that this swampland conjecture in itself has no practical consequences for an EFT theorist. 

Suppose that we have an EFT in four spacetime dimensions where
the renormalizable part of the Lagrangian respects a global $\Ztwo$ symmetry.
The SGSC states that 
this $\Ztwo$ symmetry is at best an approximate symmetry emergent in the IR, and should be broken by a higher-dimensional operator
\begin{align}
    \mathcal{L}_{\cancel{\Ztwo}}
    = \frac{1}{\Lambda_{\rm QG }^n} \mathcal{O}_{4+n} \; ,
\label{eq:breaking}
\end{align}
for some positive integers $n$. Note that in this expression we have 
included a numerical coefficient into the definition of $\Lambda_{\rm QG}$,
so that the coefficient is normalized to be exactly one. 

Now, there is a serious problem with phenomenological applications of SGSC:
we can for example satisfy the SGSC if we have a
single term of eq.~\eqref{eq:breaking}, say, for $n=100$, with $\Lambda_{\rm QG}=M_{\rm Pl}$ (with all terms with $n<100$ absent).
If such a possibility is allowed then the effect of global symmetry breaking is so tiny and has no practical consequences for model building, and we can safely disregard the SGSC.

We will therefore formulate a stronger version of the 
\emph{Refined Swampland Global Symmetry Conjecture (RSGSC)}:

\medskip

\noindent
\textbf{Refined Swampland Global Symmetry Conjecture}:
The breaking of the global symmetry, as predicted by the weak swampland
global symmetry conjecture is caused by the lowest higher-dimensional operator
consistent with the EFT (i.e.\ an operator \eqref{eq:breaking} with smallest possible value of
 positive integer $n$ allowed inside the EFT).
 
\medskip
 
In the following, we consider the leading dimension-five operator in four spacetime dimensions,
which will be present in the models to be considered below.

While we do not have a general proof for the refined version of the swampland global symmetry conjecture,
it is natural to assume this stronger version of the conjecture, at least in the absence of significant fine-tuning. 
The violation of the global symmetry can be, for example, caused by the effects of gravitational instantons (wormholes) \cite{Giddings:1987cg,Lee:1988ge,Abbott:1989jw,Coleman:1989zu}. 
Here the most typical situation is to consider a global $U(1)$ symmetry acting on a complex scalar field by a phase rotation. The phase plays the role of the axion, whose shift symmetry is broken by the effect of the gravitational instanton, 
already at the leading order ($n=5$).\footnote{Note that this statement here does not require the resolution of the axion quality problem~\cite{Kamionkowski:1992mf,Holman:1992us,Barr:1992qq,Kallosh:1995hi,Svrcek:2006yi}.), which asks for quantitative estimate of the size of the global-symmetry breaking operator.}
For a general global $\mathcal{Z}_2$ symmetry,
the precise discussion will depend on the details of the setup. We would, however, still generally expect that higher-dimensional operators of the form \eqref{eq:breaking} will always be generated by gravitational instantons
(see \cite{Daus:2020vtf} and references therein), for all values of $n$: the effects of global-symmetry violation originate from gravity, which couples universally to 
all fields. 
While there will in general be other sources of global symmetry violations, this means we will obtain a global symmetry breaking 
operator \eqref{eq:breaking} with a non-zero coefficient 
long as there is no extra fine-tuning to set the overall coefficient to be exactly zero; such a fine-tuning, however, is unlikely in the string theory landscape, which is believed to contain only finitely-many possibilities \cite{Acharya:2006zw}.

In the following, we assume RSGSC and work out the consequences of the conjecture.

Having specified the global-symmetry breaking operator,
the next question is to estimate the size of the scale $\Lambda_{\rm QG}$. 
In the literature one often encounters the statement that
we expect $\Lambda_{\rm QG} \sim \mathcal{O}(M_{\rm Pl})$~\cite{Addazi:2021xuf}.
We emphasize, however, that this is not necessarily required by QG in itself. Indeed, a global symmetry can be broken by non-perturbative instanton effects (e.g.\ D-brane instanton in string theory \cite{Blumenhagen:2006xt,Florea:2006si,Blumenhagen:2009qh} or gravitational instanton \cite{Giddings:1987cg,Lee:1988ge,Abbott:1989jw,Coleman:1989zu}),
in which case the operator in eq.~\eqref{eq:breaking} is suppressed by a factor $e^{-\mathcal{S}}$, where the dimensionless parameter $\mathcal{S}$ represents the size of the action of the non-perturbative instanton. In this case, the scale $\Lambda_{\rm QG}$ 
should be estimated as\footnote{The actual estimate of the size of the higher-dimensional operator in eq.~\eqref{eq:breaking} contains, in addition to the exponential factor $e^{-\mathcal{S}}$, 
powers of $\Lambda_{UV}/M_{\rm Pl}$
(where $\Lambda_{\rm UV}$ is the UV cutoff of the theory) and an overall numerical coefficient. These extra factors can further suppress the size of the higher-dimensional operators, and hence raise the size of $\Lambda_{\rm QG}$ (cf.\ \cite{Daus:2020vtf}). For the rest of this paper, we simplify the discussion by absorbing these factors into the definition of $\mathcal{S}$.}
\begin{align} 
    \label{Lambda_QG}
    \Lambda_{\rm QG} \sim M_{\rm Pl}\, e^{\mathcal{S}} \gg M_{\rm Pl} \; .
\end{align} 
Note that the appearance of an exponential factor in the energy scale is familiar for asymptotically free theories, where a new scale, which is exponentially suppressed from the UV cutoff scale, appears via dimensional transmutation. The appearance of the new scale in eq.~\eqref{Lambda_QG} is based on a similar non-perturbative mechanism, except that the new scale here is exponentially enhanced (instead of exponentially suppressed), since we are dealing with higher-dimensional non-renormalizable operators in the theory.

Of course, practical implications of the global symmetry breaking depend heavily on the 
exact values of $\Lambda_{\rm QG}$. We have already emphasized that this scale can be 
much higher than the Planck scale, but how large can it be?

One problem where this question has been widely discussed 
is the QCD axion, which is considered to be one of the most attractive solutions of the 
strong CP problem. Here we are invoking a global $U(1)$ Peccei-Quinn (PQ) symmetry \cite{Peccei:1977hh,
Peccei:1977ur}, and in order to solve the strong CP problem
the value of $\mathcal{S}$ should 
satisfy $\mathcal{S}\gtrsim 190$, and this results in an extremely high energy scale $\Lambda_{\rm QG}\sim 10^{100}~\textrm{GeV}$.

In principle, even if the global symmetry breaking operator \eqref{eq:breaking}
exists for the leading dimension-5 operator (with $n=1$),
its effects will be practically negligible if we can take $\Lambda_{\rm QG}$ extremely large.
We might therefore conclude that RSGSC again has no practical consequences.

One should quickly point out, however, that it is actually very difficult to 
obtain such an extremely high energy scale. 
One might, for example, imagine that an exponential factor with a large value of $\mathcal{S}$ can be generated by warped throats  \cite{Randall:1998uk} and conformal sequestering \cite{Luty:2000ec}. 
It is, however, non-trivial to realize these scenarios in string theory, see, e.g.\ \cite{Anisimov:2001zz,Anisimov:2002az,Kachru:2006em}.
Moreover, even if one manages to obtain a large value of $\mathcal{S}$ in a certain mechanism,
there are typically many competing effects to generate the global-symmetry breaking operator in eq.~\eqref{eq:breaking}, and in order to 
obtain a large value of $\Lambda_{\rm QG}$ we need to control
\emph{all} such contributions. For this reason, we need a very careful analysis which will depend on much of the details of string theory compactifications. (For the QCD axion, that we 
encounter an extremely high energy scale
which is hard to achieve in QG is known as the axion quality problem~\cite{Kamionkowski:1992mf,Holman:1992us,Barr:1992qq,Kallosh:1995hi,Svrcek:2006yi}.)

In this paper, we take a pragmatic approach and we keep the size of $\Lambda_{\rm QG}$ as a free parameter
whose value can be constrained by phenomenological and cosmological considerations.
As in our previous paper \cite{King:2023ayw},
we will discuss the value of non-perturbative action 
$\Lambda_{\rm QG}\sim (10^{20} \cdots 10^{35})~\textrm{GeV}$, which corresponds to the value $\mathcal{S}\sim (4 \cdots 38)$. One can motivate this value from a general estimate
$\mathcal{S}\sim \mathcal{O}(M_{\rm Pl}^2/\Lambda_{\rm UV}^2)$ \cite{Fichet:2019ugl,Daus:2020vtf},
where the UV cutoff $\Lambda_{\rm UV}$ of the theory is taken as $\Lambda_{\rm UV} \lesssim M_{\rm Pl}$; this is a minimalistic situation where the EFT does not require new physics up to slightly below the Planck scale.\footnote{In the estimate from the axionic version of the weak gravity conjecture, we have $\Lambda_{\rm UV}\lesssim L^{-1}:=\sqrt{f M_{Pl}}$, where $L$ is the size of the Giddings-Strominger wormhole~\cite{Giddings:1987cg} supported by an axion and $f$ is the decay constant of the axion. If we assume the saturation of the inequality, our values of $\mathcal{S}$ correspond to $f$ slightly below $M_{\rm Pl}$. We note, however, that a more precise version of the the axionic weak gravity conjecture, including precise numerical coefficients, will be needed for more precise estimate. Note also that we cannot automatically make $\mathcal{S}$ large by simply making $\Lambda_{\rm UV}$ small; larger $\mathcal{S}$ means smaller global symmetry breaking contributions from the wormhole, thereby rendering other completing effects of global symmetry violation potentially more dominant.}  Note that our parameter region requires much smaller values of the non-perturbative action $\mathcal{S}$ than the QCD axion, and hence one can regard our scenario as more ``natural'' than the QCD axion.

While the scale $\Lambda_{\rm QG}$ can be different for different higher-dimensional operators,
here we assume that all different 
$\Ztwo$ global symmetries are broken by higher dimensional operators associated with the {\it same} energy scale $\Lambda_{\rm QG}$, at least as far as those relevant for phenomenological considerations
(see discussion in \cite{King:2023ayw} for motivations of this assumption from string theory).

Let us note that the high value of the $\Lambda_{\rm QG}$ can in principle be 
imitated by non-quantum-gravity effects. For example, the QG scale \eqref{Lambda_QG}
can be imitated by starting with some scale $\Lambda_{\textrm{non-Pl}} \ll M_{\rm Pl}$
and then with a huge value of the action $\mathcal{S}'$:
\begin{align}
    \Lambda_{\textrm{non-QG}} \sim \Lambda_{\textrm{non-Pl}}\, e^{\mathcal{S}'}  \; ,
\end{align}
if the value of the action is large enough:
\begin{align}
    \mathcal{S}' = \mathcal{S}  \log(M_{\rm Pl} / \Lambda_{\textrm{non-Pl}}  ) \gg  \mathcal{S} \; .
\end{align}

This would be, however, more difficult to achieve than the QG scenarios, since we need a much larger value of 
$\mathcal{S}'$, and hence much more careful control of all the non-perturbative corrections in the theory.

\section{Simplified Model for BSM Scenarios} \label{sec:simplified}

In this work, we augment the Standard Model particle spectrum with one scalar field $S$ and one fermion $\chi$, 
either of which is a singlet under the SM gauge symmetry. Furthermore, we invoke additional discrete symmetries, $\mathcal{Z}_2^{\text{DW}}\times \mathcal{Z}_2^{\text{DM}}$, under which all SM fields transform trivially whereas the newly introduced particles carry an odd charge under their respective discrete symmetry groups. 

The role of the DM in this particular setup is played by the fermion $\chi$  with mass term $m_{\rm DM
}\overline{\chi^c}\chi/2$. The $\mathcal{Z}_2^{\text{DW}}\times \mathcal{Z}_2^{\text{DM}}$-conserving dimension-five Lagrangian involving the DM and other fields can be written as 
\begin{eqnarray}
   \mathcal{L}  &=& \frac{ 1}{\Lambda_{\text{FI}}}  \overline{\chi}\chi S^2+ \frac{ 1}{\Lambda_{\text{FI}}}  \overline{\chi}\chi H^\dagger H \; ,
    \label{Lag}
\end{eqnarray}
where $\Lambda_{\text{FI}}$ denotes the effective freeze-in scale and $H$ is the SM Higgs doublet field. The most general scalar potential involving different scalars at the tree level can be expressed as
\begin{eqnarray}
    V&=& \mu ^2  H^\dagger H+ \lambda (H^\dagger H)^2 +\lambda _{hs} H^\dagger H S^2  \nonumber \\ && 
    +\frac{\lambda_s}{4} (S^2-v_s^2)^2   \; .
    \label{eq:potential}
\end{eqnarray}
The scalar $S$ can acquire a VEV $v^{}_s$ in the early Universe, then the $\Ztwo^{\text{DW}}$ symmetry is spontaneously broken, generating DWs in the process. Subsequent to the EW symmetry breaking, the CP-even component of the Higgs doublet can mix with the scalar $S$. Nevertheless, we work in the limit where $v_s \gg v_h$ (with $v_h$ being the vev of the neutral component of $H$), and we also assume the coupling $\lambda^{}_{hs}$ to be sufficiently small. These assumptions imply that the mixing angle between $S$ and the CP-even component of $H$ is negligibly small~\cite{Bhattacharya:2019tqq}. The scalar potential on the other hand should be bounded from below to make the electroweak vacuum stable. This poses additional constraints on the scalar couplings which can be found in ref.~\cite{Bhattacharya:2019tqq}. 

Following the RSGSC, the $\ZDW$ symmetry is broken by higher-dimensional operators of the form 
    \begin{eqnarray}
        \Delta V  &=& \frac{1}{\Lambda_{\rm QG}} (\alpha_{1}  S^5 + \alpha_{2}  S^3 H^2 + \alpha_{3}  S H^4 ) \; , 
        \label{eq:bias}
    \end{eqnarray}
whereas both $\ZDW$ and $\ZDM$ symmetries are broken explicitly by,  
\begin{eqnarray}
        \mathcal{L}_{\text{break}}  &=& \sum_{i=1,2,3} \frac{ \beta_i}{\Lambda_{\rm QG}}  S\overline{\ell_{L_i}}\widetilde{H}\chi  \; ,
        \label{eq:Z2both}
    \end{eqnarray}   
where $\widetilde{H}:=i\sigma_2 H^*$. We assume that all the global symmetries are broken by a common scale, as explained in \cref{sec:QG}.
This means that we simply take all the dimensionless coefficients $\alpha_i, \beta_i$ in eqs.~\eqref{eq:bias} and \eqref{eq:Z2both} to be of the same order, and we can make all of them to be ${\cal O}(1)$ by redefining $\Lambda_{\rm QG}$. 

After the breaking of EW symmetry,  eq. \eqref{eq:Z2both} results in, 
\begin{eqnarray}
        \mathcal{L}_{\text{break}}  &=& \sum_{i=1,2,3} \frac{ \beta_iv_s v_h}{\sqrt{2}\Lambda_{\rm QG}}  \overline{\nu_{L_i}}\chi  \; .
        \label{eq:Z2both2}
    \end{eqnarray} The above equation generates a mass term for the SM neutrinos which is similar to the Dirac mass obtained in the Type-I seesaw mechanism \cite{Minkowski:1977sc,Yanagida:1979as,Yanagida:1979gs,GellMann:1980vs,Mohapatra:1979ia,Schechter:1980gr,Schechter:1981cv} of neutrino mass generation and hence we can write,
    
    \begin{equation}
        m_{D_i}= \frac{ \beta_i v_s v_h}{\sqrt{2}\Lambda_{\rm QG}}.
    \end{equation}
As a result of eq. \eqref{eq:Z2both2}, the SM neutrino also mixes with the DM $\chi$. The mixing angle in this situation can be expressed analogous to the active-sterile mixing obtained after diagonalizing the neutrino mass matrix in the Type-I seesaw mechanism. Following \cite{Datta:2021elq}, the mixing angle $\theta$ can be written as 
\begin{equation}
\theta
 \simeq\sum_{i=1,2,3}\left(\frac{m_{D_i}}{m_{\text{DM}}}\right)
 =\sum_{i=1,2,3}\left(\frac{\beta_i v_sv_h}{\sqrt{2}\Lambda_{\text{QG}}m_{\text{DM}}} \right)\; .   
 \label{eq:mixingangle}
\end{equation}
The DM-neutrino mixing allows the DM to decay to the SM particles. The late-time decay of the DM spoils the success of the big-bang nucleosynthesis (BBN), which in turn sets a lower bound on the DM lifetime. As we discussed in \cite{King:2023ayw}, though one often encounters a large number $\mathcal{O}(100)$ of SM-singlet moduli fields in string theory compactifications and we are considering a simplified model with only one scalar and one fermion, it is not oversimplified and does capture the qualitative features of a large class of string-inspired models involving many moduli fields, since the DM abundance is actually dominated by the contribution of a single particle species, and the GW spectrum from the DWs annihilation is also typically dominated by the contribution of a single scalar field with the largest VEV. 

\section{DM phenomenology}\label{sec:DM}

Although the particle nature of the DM is still a puzzle, one prefers a scalar DM over a fermionic DM in a WIMP scenario. 
A scalar DM (singlet under the SM gauge symmetry) has a quartic interaction\footnote{If the scalar DM is considered to be a part of a multiplet, it can have additional gauge interactions that can keep it in equilibrium in the early Universe. The same interaction can contribute strongly towards its annihilation.} with the SM Higgs boson that helps it to be in equilibrium with the thermal bath in the early Universe. The same interaction also forces DM to annihilate into the SM particles via Higgs mediation and leads to their freeze-out. 

The situation is very different if we replace a scalar with a fermion: a fermion (especially
if it is considered to be a singlet under the SM gauge symmetry) has several issues as a WIMP candidate. 
To start with, if protected by a symmetry (for example a $\mathcal{Z}_2$ symmetry), the DM will remain completely decoupled and hence absolutely stable. Under such a situation, the DM cannot be produced by any interaction (except the gravitational one). If one relaxes the symmetry argument and allows the DM to couple with the SM particle, the only renormalizable interaction it can have is the Yukawa interaction with the SM Higgs and lepton doublet. Once the EW symmetry is spontaneously broken, such interactions allow the DM to mix with the SM neutrinos and as a result, the DM can decay. Now, in order for the DM to be in equilibrium in the early Universe, the mixing angle has to be very large. Such a large mixing angle is strictly prohibited by several cosmological experiments as it leads to the faster decay of the DM. 

This situation can still be avoided if a DM is charged under a bigger symmetry, e.g. if it is charged under the $U(1)_{B-L}$ symmetry ~\cite{Bhattacharya:2019tqq}. Here the DM can achieve equilibrium with the SM bath in the early Universe but it is found that the fermionic DM can only provide the correct DM abundance in near resonance regions in this situation. Hence, the allowed DM parameter space is highly restrictive. 

Although a WIMP has been an attractive DM candidate, its presence has been heavily scrutinized in recent times. The tight constraints from direct and indirect search experiments on the WIMP parameter space have motivated the particle-physics community to explore different DM paradigms. One such paradigm is where the DM is produced slowly from the thermal plasma as a result of its feeble interaction with the SM bath. Here, the initial abundance of the DM is considered to be zero and, as the temperature of the Universe falls, the DM is expected to be dominantly produced by the decay or scattering of other particles. Due to the involvement of the tiny strengths of the couplings at play here, the interaction rate(s) of the DM with the SM bath is always smaller than the Hubble expansion rate. The freeze-in production of DM dominantly takes place when the temperature of the thermal soup is of the order of the mass of the mother particle responsible for the production of DM.

A fermion as a FIMP DM has always attracted the attention of DM community. Due to its limited number of interactions, a fermion, if considered as a DM, finds it difficult to enter into the thermal equilibrium with the SM bath (unless it interacts very strongly with the bath). Hence, it is more convenient to assume that it interacts feebly and never equilibrates with thermal plasma. Motivated by this, in the present setup we consider a fermion DM $\chi$ which is produced slowly from the thermal bath via scattering as a result of interactions shown in eq.~\eqref{Lag}. Due to the involvement of non-renormalizable operators, the production of DM proceeds through UV freeze-in and is sensitive to high temperatures. The feeble interaction and out-of-equilibrium production of the DM is guaranteed by the choice of the freeze-in scale, $\Lambda_{\text{FI}}$. One can write the DM abundance produced in a UV freeze-in scenario as~\cite{Hall:2009bx,Elahi:2014fsa}
\begin{equation}
    Y_{\text{DM}}^{\text{UV}}\simeq 3\frac{180}{1.66\times (2\pi)^7g^S_\star\sqrt{g_\star^\rho}}\bigg(\frac{T_{\text{RH}}M_{\text{Pl}}}{\Lambda_{\text{FI}}^2}\bigg) \; ,
\end{equation}
where $M_{\text{Pl}}=1.2\times10^{19}$ GeV and $g^{S,\rho}_\star$ is the effective number of degrees of freedom in the bath.  In a UV freeze-in scenario, the highest temperature of the Universe is considered to be the reheating temperature, $T_{\text{RH}}$. Typically, the lower bound on $T_{\text{RH}}$ comes from the measurement of light element abundance during BBN, which requires $T_{\text{RH}}>\mathcal{O}(\text{MeV})$~\cite{Giudice:2000ex}. The upper bound, on the other hand, may emerge from (i) cosmological gravitino problem~\cite{Kawasaki:2004qu} in the context of a supersymmetric framework that demands $T_{\text{RH}}\leq 10^{10}$ GeV to prohibit thermal gravitino overproduction and (ii) simple inflationary scenarios that require at most $T_{\text{RH}}\sim 10^{16}$ GeV~\cite{Linde:1990flp,Kofman:1997yn} for successful inflation. In view of this, the reheating temperature of the Universe can thus be treated as a free parameter.

\begin{figure}[t!]
  \centering
  \includegraphics[width=0.96\linewidth]{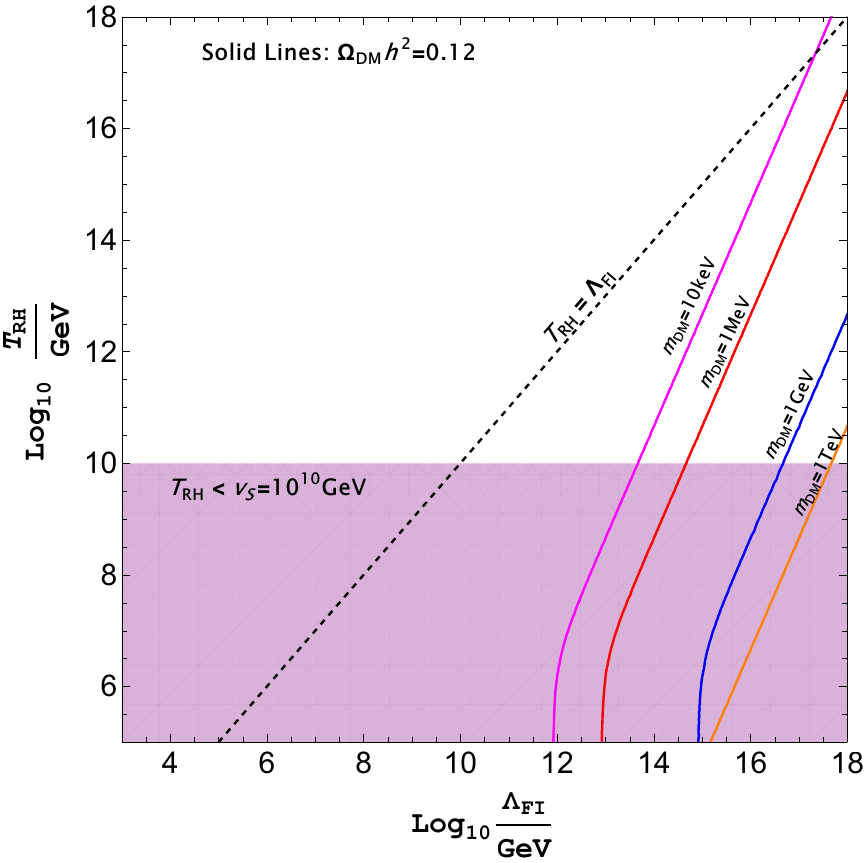}  
  \caption{Contours satisfying the correct magnitude of DM relic density for different DM masses are shown by solid lines in the $T_{\text{RH}}$--$\Lambda_{\text{FI}}$ plane. Yhe black dashed line corresponds to $T_{\text{RH}}=\Lambda_{\text{FI}}$, and the purple shaded region corresponds to parameter space where $T_{\text{RH}}<v_s$. Here we set $v_s=10^{10}$ GeV and $m_s=1$ TeV.}
  \label{fig:summary1}
\end{figure}

On the other hand, the present set-up also provides an option of producing the DM via IR freeze-in. Once the symmetries such as the global $\ZDW$ or the EW symmetry are broken, the DM can be produced from the decay of the two scalars ($s$ and $h$) via the renormalizable operators like $v_s s 
\overline{\chi}\chi/\Lambda_{\text{FI}}$ or $v_h h\overline{\chi}\chi/\Lambda_{\text{FI}}$ following from eq.~\eqref{Lag}. In such a situation the DM abundance can be estimated using~\cite{Barman:2020plp}
\begin{equation}
   Y_{\text{DM}}^{\text{IR}}\simeq \frac{135\, M_{\text{Pl}}}{1.66\times 8\pi^3g^S_\star\sqrt{g_\star^\rho}}\bigg(\frac{\Gamma_{s\to\chi\chi}}{m_s^2}+\frac{\Gamma_{h\to\chi\chi}}{m_h^2}\bigg) \; . 
\end{equation}

If the breaking scales are high or comparable to the reheating temperature one cannot ignore the contribution of IR freeze-in in the present setup, under which the total DM abundance can be expressed as 
\begin{equation}
   Y_{\text{DM}}^{\text{Total}}= Y_{\text{DM}}^{\text{UV}}+Y_{\text{DM}}^{\text{IR}} \; . 
\end{equation}
Following this one can obtain the DM relic density at the present epoch using
\begin{equation}
    \Omega_{\text{DM}} h^2=2.75\times 10^8 \bigg(\frac{m_{\text{DM}}}{\text{GeV}}\bigg) Y_{\text{DM}}^{\text{Total}} \; .
\end{equation}

In Fig.~\ref{fig:summary1}, we display the contours (solid lines) in the $T_{\text{RH}}$--$\Lambda_{\text{FI}}$ plane that satisfies the observed DM abundance in the Universe, i.e. $\Omega_{\text{DM}} h^2\simeq0.12$, for different DM masses. In this plot, we have set  $v_s=10^{10}$ GeV and $m_s=1$ TeV for the demonstration purpose.  It should be noticed that an identical behavior in the pattern is observed for the DM masses which are below the mass of the scalar $m_s=1$ TeV. Here for a low reheating temperature, the DM relic density is predominantly obtained by the IR freeze-in and hence the solid lines are independent of $T_{\text{RH}}$. For a $T_{\text{RH}}$ which is comparable or larger than  the $v_s$, a clear variation of $T_{\text{RH}}$ with $\Lambda_{\text{FI}}$ is observed. As expected, in this regime, the DM abundance is dictated by the UV freeze-in and a large $T_{\text{RH}}$ is required for a larger value of  $\Lambda_{\text{FI}}$ to satisfy the observed DM abundance. However, for a DM with mass $m_\text{DM}=1$ TeV, its production from the decay of the $s$ is kinematically disallowed and IR freeze-in would not take place, hence no bending is observed in the solid orange line. We also demarcate the regions where $T_{\text{RH}}=\Lambda_{\text{FI}}$ shown by black dashed line and  $T_{\text{RH}}< v_s$ in purple. The purple region is disallowed as in this region the $\ZDW$ breaking scale lies above the reheating temperature.

\section{Indirect detection of QG scale}\label{sec:indirect}

As mentioned in the previous sections, the fermionic DM $\chi$ in our framework can have Yukawa interactions [eq. \eqref{eq:Z2both}] with the SM neutrinos, which allow $\chi$ to decay into SM particles, thus providing us with possible means to indirectly detect DM.

One approach to place stringent constraints on the decaying DM is via the detection of the galactic and extra-galactic diffuse X/$\gamma$-ray background. As a result of mixing discussed in eq. \eqref{eq:mixingangle}, the DM $\chi$ can now interact with the SM $W^\pm$ and $Z$ bosons through charge-current and neutral-current interactions. If kinematically allowed, these interactions can lead to a dominant two-body decay at one-loop level mediated via $W^\pm$ and SM charged leptons, and three body decays of $\chi$ at tree-level (mediated via $W^\pm$). At one-loop level $\chi$ can decay to photons and active neutrinos via $\chi \to \nu \gamma$, the decay rate of which is given by~\cite{Shrock:1982sc,Essig:2013goa}
\begin{equation}
\begin{split}
\tau_{\chi \rightarrow \nu \gamma} & \simeq\left(\frac{9 \alpha_{\mathrm{EM}} \sin ^2 \theta}{1024 \pi^4} G_F^2 m_{\rm DM}^5\right)^{-1} \\
& \simeq 1.8 \times 10^{17}~{\rm s}\left(\frac{10 \mathrm{MeV}}{m_{\rm DM}}\right)^5\left(\frac{\sin \theta}{10^{-8}}\right)^{-2} \; ,
\end{split}
\label{eq:lifetime}
\end{equation}
with $\alpha_{\rm EM} = 1/137$ being the electromagnetic fine-structure constant and $G_F$ denoting the Fermi constant. Apart from the above two-body radiative decay, the three-body decay channel $\chi \to e^+ e^- \nu$, the decay rate of which can be approximately expressed as \cite{Ruchayskiy:2011aa,Essig:2013goa}
\begin{equation}
\begin{split}
    \tau_{\chi \to e^+ e^- \nu} & \simeq \left(   \frac{c_\alpha \sin^2 \theta}{96 \pi^3} G^2_F m^5_{\rm DM}\right)^{-1} \\
    & \simeq 2.4 \times 10^{15}~ {\rm s} \left(\frac{10 \mathrm{MeV}}{m_{\rm DM}}\right)^5\left(\frac{\sin \theta}{10^{-8}}\right)^{-2} \;, 
    \end{split}
    \label{eq:lifetime3}
\end{equation}
where $c_{\alpha}:=(1+ 4 \sin^2 \theta_W + 8 \sin^4 \theta_W)/4 \simeq 0.59$ with $\theta_W$ being the weak mixing angle.
Contributions from the above two channels to the X/$\gamma$-ray fluxes are roughly at similar levels~\cite{Essig:2013goa}. The diffuse X/$\gamma$-ray background is actively searched by numerous projects. Null observation of the expected spectral line sets a lower bound on the lifetime of DM, which can then be converted to the constraints on the DM mass $m_{\rm DM} $ and the mixing angle $\theta$. The observed diffuse photon spectra data obtained from the HEAO-1~\cite{Gruber:1999yr}, INTEGRAL~\cite{Bouchet:2008rp}, COMPTEL~\cite{Sreekumar:1997yg} and EGRET~\cite{Strong:2003ey} satellites can restrict the parameter space of $m_{\rm DM}$ and $\theta$ within the range $0.01~{\rm MeV} \lesssim m_{\rm DM} \lesssim 100~{\rm MeV}$, which can be parametrized as $\theta^2 \lesssim 2.8 \times 10^{-18}({\rm MeV}/m_{\rm DM})^5$~\cite{Boyarsky:2009ix}. With  the help of the expression of $\theta$ in eq.~\eqref{eq:mixingangle}, we have
\begin{equation}
\left(\frac{3 v_sv_h}{\sqrt{2}\Lambda_{\text{QG}}m_{\text{DM}}}\right)^2 \lesssim 2.8\times10^{-18}\left(\frac{\text{MeV}}{m_{\text{DM}}}\right)^5 \; ,
\label{DM-decay}
\end{equation}
where we consider that the DM interacts with different generations of neutrinos with the identical ${\cal O}(1)$  strength, i.e. $\beta_1=\beta_2=\beta_3=1$. On the other hand, the Fermi Large Area Telescope (Fermi-LAT)~\cite{Fermi-LAT:2012pls,Fermi-LAT:2012ugx} has presented a dedicated line search of the diffuse $\gamma$-ray background, the null result of which also constrains the lifetime of the DM within the mass range $1~{\rm GeV} \lesssim m_{\rm DM} \lesssim 1~{\rm TeV}$.  Here we take the 95\% C.L. limit on the lifetime of decaying DM given by ref.~\cite{Fermi-LAT:2015kyq}, which assumes a Navarro-Frenk-White profile for the DM distribution~\cite{Navarro:1995iw}. Then we can also derive constraints on the $\{v_s, m_{\rm DM}\}$ space for given $\Lambda_{\rm QG}$ by considering the two-body and three-body decays of $\chi$.

\begin{figure*}[t!]
  \centering
  \includegraphics[width=1\linewidth]{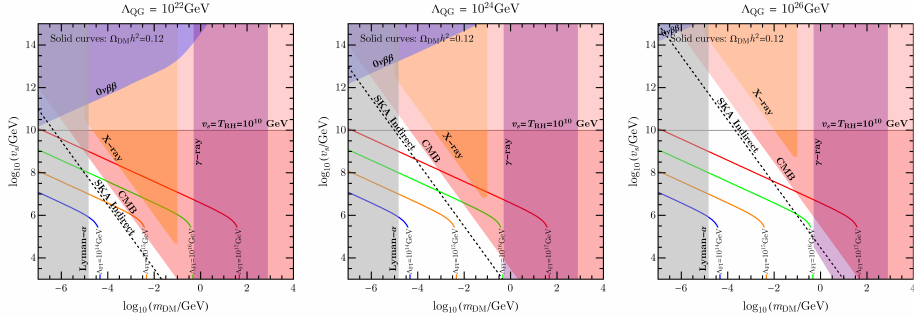}  \caption{Constraints on the DM mass $m_{\rm DM}$ and the VEV $v_s$ of the scalar field from the DM relic density and indirect detections with different values of QG scales $\Lambda_{\rm QG} = 10^{22}, 10^{24}, 10^{26}~{\rm GeV}$. The blue, orange, green, and red solid curves denote the correlations between $m_{\rm DM}$ and $v_s$ we should have to satisfy the correct DM relic density $\Omega_{\rm DM}h^2 = 0.12$ for different freeze-in scales $\Lambda_{\rm FI} = 10^{14}, 10^{15}, 10^{16}, 10^{17}~{\rm GeV}$, where we set $T_\text{RH}=10^{10}$ GeV and $m_s=1\,{\rm TeV}$. The blue-, orange-, purple-, red-, and gray-shaded areas represent respectively the excluded regions by the $0\nu\beta\beta$ decay, X-ray, $\gamma$-ray, CMB, and Lyman-$§alpha$ observations. \textbf{(Note that in this plot $\lambda_S$ is varying.)}}
  \label{fig:summary2}
\end{figure*}

If the decay of DM occurs during or after the recombination, it can leave an imprint on the CMB power spectrum. To be specific, the injected energy due to the DM decay will lead to the reionization of the intergalactic medium, thus modifying the CMB power spectrum.  Accurate measurements of the CMB spectrum have been implemented by recent experiments including WMAP~\cite{WMAP:2012nax}, ACT~\cite{ACTPol:2014pbf}, SPT~\cite{Hou:2012xq} and Planck~\cite{Planck:2015fie}. The 95\% C.L. lower bounds on the DM decay lifetime have been gained in ref.~\cite{Slatyer:2016qyl}, showing that $\tau_{\rm DM} \gtrsim 10^{25}~{\rm s}$ for decays into both $e^+e^-$ pairs and photons.

The decay of DM may also induce radio signals originating inside the DM-dominated galaxies and clusters, if the produced $e^+e^-$ pairs undergo energy loss via electromagnetic interactions in the interstellar medium. Such radio waves can be tested by radio telescopes like the Square Kilometer Array (SKA) radio telescope~\cite{Colafrancesco:2015ola}. It is found that the DM decay width up to $\Gamma_\text{DM}\gtrsim10^{-30}~{\rm s}^{-1}$ is detectable at SKA assuming  100-hour observation time~\cite{Dutta:2022wuc}.

Moreover, if the neutrino is of Majorana-type, it could serve as the intermediate particle in the processes of neutrinoless double-beta (0$\nu\beta\beta$) decays, where the heavy sterile neutrino may also play a role due to the active-sterile mixing. Therefore 0$\nu\beta\beta$ decay experiments are capable of setting stringent constraints on active-sterile neutrino mixing in the electron flavor, especially when the sterile neutrino is in the small mass region~\cite{Bolton:2019pcu, Atre:2009rg, Deppisch:2015qwa,Huang:2019qvq}. Here we work in the scenario that is analogous to the canonical Type-I seesaw mechanism, whereas we assume other sterile neutrinos except $\chi$ are sufficiently heavy. As a result, their mixing with the electron neutrino should be negligibly small and irrelevant to our discussion. Here we use the upper limits on the active-sterile mixing in which the $0\nu\beta\beta$ decay is driven by a single sterile neutrino (c.f. the red curve in Fig.~12 of ref.~\cite{Bolton:2019pcu}), and translate them into the constraints on the $v_s-m_{\rm DM}$ plane using eq.~(\ref{eq:mixingangle}).

The Lyman-$\alpha$ forest serves as a measurement of the hydrogen clouds in the line of sight through the absorption lines of the distant quasars~\cite{Viel:2005qj,Viel:2013fqw,Palanque-Delabrouille:2019iyz,Garzilli:2019qki}. Its observation sheds light on the intermediate to small structure formation. DM with large momentum will in general have a large free streaming length that can lead to the suppression in the formation of the small-scale structures, which provides the constraint on the DM mass. A recent study ~\cite{Decant:2021mhj} has shown that Lyman-$\alpha$ bound can exclude DM mass with $\lesssim \text{15 keV} $ if the DM is produced through a freeze-in mechanism.

In \cref{fig:summary2}, we present the aforementioned constraints on $v_s$ and $m_{\rm DM}$ with varying $\Lambda_{\rm QG}$ from CMB, X/$\gamma$-ray and $0\nu\beta\beta$ decay observations, together with the region that can be tested by SKA in the future. One could observe that constraints from different aspects squeeze the allowed parameter space to be roughly a triangle region, the area of which becomes smaller as  $\Lambda_{\rm QG}$ decreases. Observations on CMB and $\gamma$-ray background strongly disfavor DM with $m_{\rm DM} \gtrsim 100~{\rm MeV}$, especially for small $\Lambda_{\rm QG}$ and large $v_s$. It is interesting to note that the allowed parameter space should further satisfy the correct DM relic density. As can be seen from the solid colored curves in Fig.~\ref{fig:summary2}, in order to generate the correct DM relic density in our current Universe, $v_s$ and $m_{\rm DM}$ should satisfy certain relations depending on the values of $\Lambda_{\rm FI}$. For $T_{\rm RH} = 10^{10}~{\rm GeV}$ we choose, the UV and IR freeze-in can be comparable when $v_s \gtrsim 10^{5}~{\rm GeV}$. Nevertheless, if $v_s$ drops below $10^{5}~{\rm GeV}$, the UV freeze-in becomes dominant, which explains why these curves will finally fall into several certain values of $m_{\rm DM}$. In addition, if a larger $\Lambda_{\rm FI}$ is taken into account, the values of $v_s$ and $m_{\rm DM}$ required to achieve the correct relic density would also be greater, facing more stringent restrictions from CMB and X/$\gamma$-ray observations. Therefore, the energy scales $\Lambda_{\rm FI}$ and $\Lambda_{\rm QG}$ are in fact mutually constrained. Notice that in these plots we restrict our self in the region of parameter space where $v_s\gtrsim 100$ GeV, which is because we want to have a hierarchy in between the two VEVs as also discussed in section \ref{sec:simplified}.   

\section{GWs from DWs annihilation}\label{sec:DW}

As we pointed out in our previous work \cite{King:2023ayw}, it is natural to consider the possibility that some of the many moduli in string theory result in the spontaneous breaking of a $\ZDW$ symmetry. This would lead to topological defects known as DWs which would result in different parts of the Universe being patches of different vacua~\cite{Kibble:1976sj}.
If such DWs were absolutely stable, they would come to dominate the Universe and conflict with the observation as they dilute linearly with the scale factor---much faster than radiation or even matter. If they are metastable but sufficiently long-lived until the beginning of BBN, they can also be inconsistent with current observations~\cite{Zeldovich:1974uw}.
Fortunately, QG effects will generate explicitly $\ZDW$ breaking terms of dimension five. After replacing fields with their individual VEVs, this leads to a bias contribution to the effective potential of the form
\begin{equation}
    V^{}_{\rm bias} \simeq \frac{1}{\Lambda^{}_{\rm QG}}\left( v^5_s + \frac{v^3_s v^2_{h}}{2} + \frac{v^{}_s v^4_{h}}{4} \right) \; .
    \label{eq:model-bias}
\end{equation}
Such a bias term allows the DWs to decay which results in a characteristic stochastic GW background (SGWB) signal. Typically, $v_s \gg v_{h}$ in our analysis so there will generally be a large hierarchy amongst these operators and we can approximate
\begin{equation}
    V^{}_{\rm bias} \simeq \frac{v^5_s}{\Lambda^{}_{\rm QG}}   \; .
    \label{eq:model-bias1}
\end{equation}

Following the analysis of DW annihilation in refs.~\cite{Vilenkin:1981zs, Gelmini:1988sf, Larsson:1996sp, Hiramatsu:2013qaa, Hiramatsu:2012sc, Saikawa:2017hiv, Kitajima:2023cek,Bhattacharya:2023kws,Roshan:2024qnv}, the peak frequency and amplitude of the resulting spectrum are, respectively~\cite{Saikawa:2017hiv, Chen:2020wvu}
\begin{equation}
\begin{split}
f_p^{} & \simeq 3.75 \times 10^{-9}_{}~{\rm Hz} ~C^{-1/2}_{\rm ann} \mathcal{A}^{-1/2}_{} \widehat{\sigma}^{-1/2}_{} \widehat{V}^{1/2}_{\rm bias} \; , \\ 
\Omega^{}_{p} h^2_{} & \simeq 5.3 \times 10^{-20}_{}~ \widetilde{\epsilon} \mathcal{A}^4_{} C^2_{\rm ann} \widehat{\sigma}^{4}_{} \widehat{V}^{-2}_{\rm bias}  \; ,
\end{split}
\label{eq:peak-a}
\end{equation}
with $\widehat{\sigma}: = \sigma/{\rm TeV}^3$ ($\sigma$ is the surface tension) and $\widehat{V}_{\rm bias}: = V_{\rm bias}/{\rm MeV^4}$. In our present work, we consider a static planar domain wall lying perpendicular to the $z$-axis. Hence for the scalar potential of $S$ shown in eq.~(\ref{eq:potential}), we can obtain $s(z) = v_s {\rm tanh}(\sqrt{\lambda_s/2}v_s z)$ by solving the field equation. Then the surface tension energy density can be evaluated by integrating over the 00-component of the energy momentum tensor as
\begin{equation}
    \sigma = \int_{-\infty}^\infty {\rm d}z \left( \frac{{\rm d}s(z)}{{\rm d}z} \right) = \sqrt{\frac{8\lambda_s}{9}}v^3_s \; .
\end{equation}
In addition, we use the values of $\mathcal{A}=0.8$ for the area parameter and $C_{\rm ann}=2$ for the dimensionless annihilation constant. Furthermore, the efficiency parameter is taken to be $\widetilde{\epsilon} \simeq 0.7$ and we take it to be constant~\cite{Hiramatsu:2012sc}. Finally, we track the SM prediction for the effective number of degrees of freedom for energy and entropy density. It should be noticed that the decay products of DWs may still ruin light elements created at the epoch of BBN, even if they are annihilated before dominating the energy density of our universe. The ratio of the energy density of DWs to the entropy density at this epoch approximates to~\cite{Saikawa:2017hiv} 
\begin{equation}
    \frac{\rho_{\rm DM}}{s}(t) \simeq 2.24 \times 10^{-7}{\rm GeV} {\cal A} \widehat{\sigma} \left(\frac{t}{1\,{\rm s}}\right)^{1/2} \; .
\end{equation}
Assuming DWs significantly decay into energetic particles, their lifetime $\tau_{\rm DW}$ should be smaller than $t_{\rm ann}\lesssim 0.01\,{\rm s}$ (with $t_{\rm ann}$ being the annihilation time of the DWs) according to the constraints on the BBN energy injection~\cite{Kawasaki:2004yh, Kawasaki:2004qu}. This results in a lower bound on the bias term $V^{1/4}_{\rm bias} \gtrsim 5.07\times10^{-4}_{}~{\rm GeV}~C^{1/4}_{\rm ann}\mathcal{A}^{1/4}_{}\widehat{\sigma}^{1/4}_{}$ which corresponds to a constraint on the peak frequency $f^{}_{p} \gtrsim 0.964\times 10^{-9}~{\rm Hz}$. In any case, this is approximately the lowest frequency probed by pulsar timing arrays. 

\section{Combined constraints on the QG scale}\label{sec:combined}

\begin{figure*}[t!]
  \centering
  \includegraphics[width=1\linewidth]{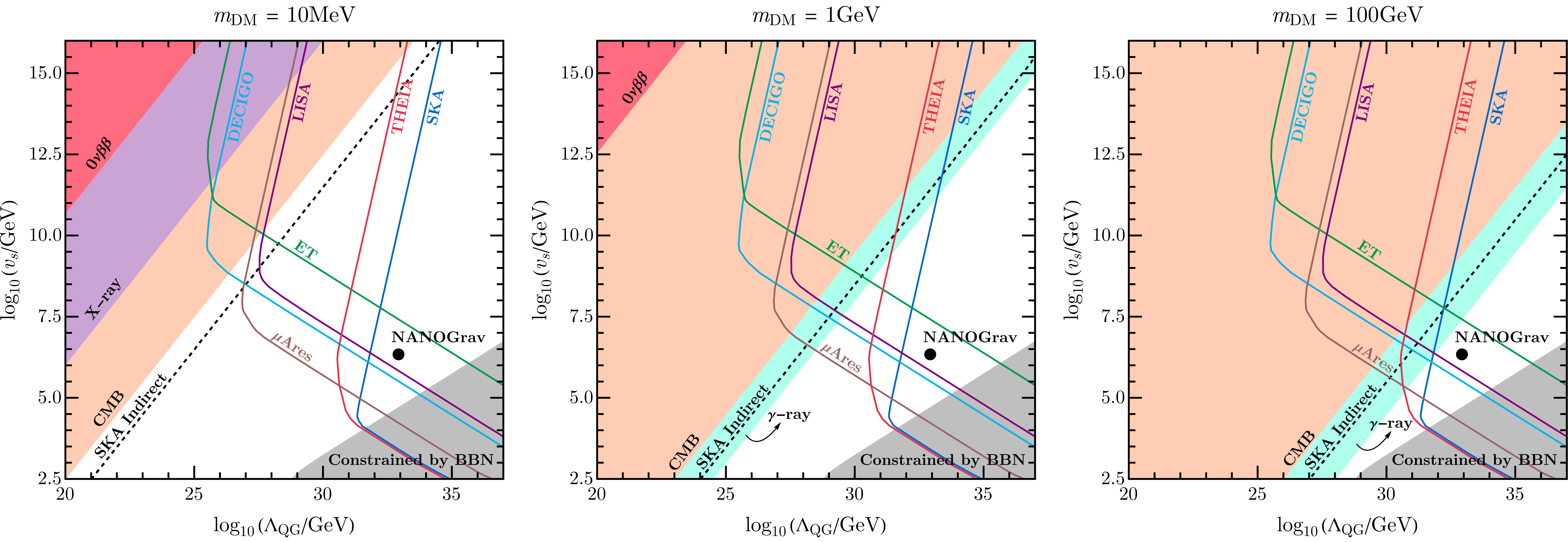}  \caption{Combined constraints on $\Lambda_{\rm QG}$ and $v_s$ from indirect DM detections and GWs observations with varying DM mass $m_{\rm DM} = 10~{\rm MeV}, 1~{\rm GeV}, 100~{\rm GeV}$. The red-, purple-, cyan- and yellow-shaded areas denote respectively the excluded regions by the $0\nu\beta\beta$ decay, X-ray, $\gamma$-ray, and CMB observations. The black dashed lines label the indirect testing capabilities of the upcoming SKA telescope. The black dots correspond to the values of $\Lambda_{\rm QG}$ and $v_s$ giving rise to $f_p = 4.07 \times 10^{-8}~{\rm Hz}$ and $\Omega_p h^2 = 1.76\times10^{-7}$ with which the SGWB is consistent with NANOGrav 15-year result, where $\lambda_s = 10^{-3}$ is assumed. The red, blue, brown, cyan, purple and green curves present the testing capabilities for the power-law SW spectra of THEIA, SKA, $\mu$Ares, DECIGO, LISA and ET with a threshold SNR $\varrho=10$. The gray-shaded regions are excluded by the requirement $\tau_{\rm DW} \lesssim 0.01~{\rm s}$.}
  \label{fig:summary}
\end{figure*}

Since QG effects motivate a common scale of DM decays and GWs from the annihilation of DWs, it is very promising that the DM indirect detections and GWs observations will provide us with combined constraints on the QG scale $\Lambda_{\rm QG}$. In \cref{fig:summary}, we summarize the constraints on $\Lambda_{\rm QG}$ and $v_s$ where different values of the DM mass $m_{\rm DM} = 10~{\rm MeV}$, $1~{\rm GeV}$ and $100~{\rm GeV}$ are chosen for illustration. The red-, purple-, cyan- and yellow-shaded regions in \cref{fig:summary} are excluded respectively by $0\nu\beta\beta$ decay, X-ray, $\gamma$-ray, and CMB observations, whereas the black dashed lines label the regions that can be potentially tested by the upcoming SKA telescope. One can clearly see which aspects of experimental constraints will be effective for a given DM mass. For small $m_{\rm DM}$, e.g. $m_{\rm DM} = 10~{\rm MeV}$, constraints from the CMB observations, the diffuse X-ray background and the $0\nu\beta\beta$ decays will come into play, where the CMB observations provide the most stringent constraint. When $m_{\rm DM}$ becomes larger ($m_{\rm DM} \gtrsim 1~{\rm GeV})$, the diffuse $\gamma$-ray background will play a crucial role in restricting the parameter space of $\Lambda_{\rm QG}$ and $v_s$. Considering the above experimental constraints collectively, we can find that smaller values of $\Lambda_{\rm QG}$ and larger values of $v_s$ tend to be favored for exclusion.

\begin{figure}[t!]
  \centering
  \includegraphics[width=1\linewidth]{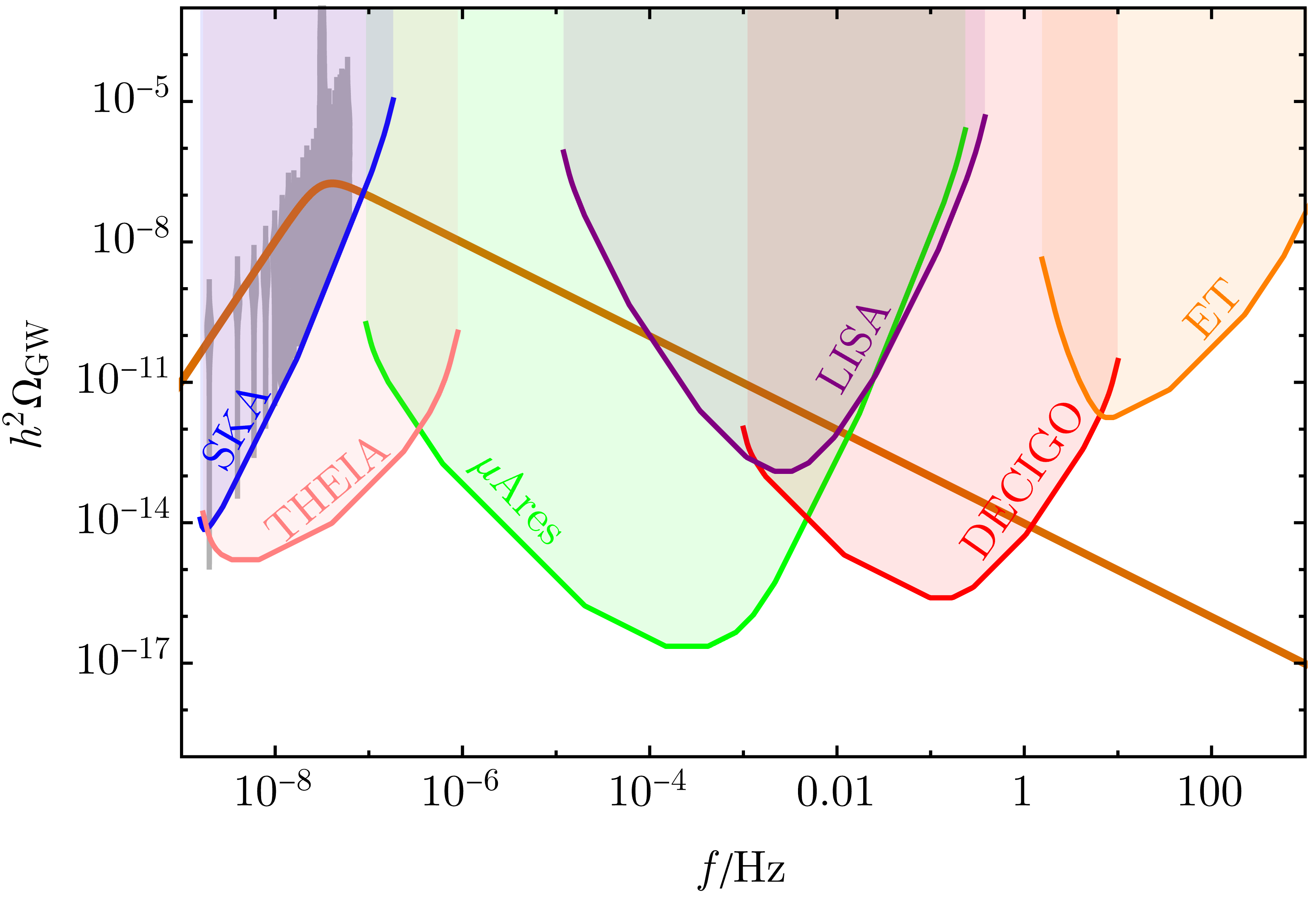}  \caption{The orange curve denotes our benchmark GW spectrum and the gray violins label the GW spectrum obtained by the  NANOGrav collaboration~\cite{NANOGrav:2023gor, NANOGrav:2023hvm}. In addition, we also show the power-law-integrated sensitivity curves of other GW detectors.}
  \label{fig:fit-DW2}
\end{figure}

Recently, the NANOGrav and several other pulsar timing array collaborations reported the evidence for a SGWB signal with the frequency around $10^{-8}~{\rm Hz}$, which indicates that the peak frequency of the GW induced by the DWs annihilation in our work may also be within this range. In this regard, we consider a benchmark with $\Lambda_{\rm QG} = 8.68 \times 10^{32}~{\rm GeV}$ and $v_s = 2.21 \times 10^6~{\rm GeV} $ which can give rise to $f_p = 4.07 \times 10^{-8}~{\rm Hz}$ and $\Omega_p h^2 = 1.76\times10^{-7}$ consistent with NANOGrav 15-year result~\cite{NANOGrav:2023gor, NANOGrav:2023hvm}. We label this benchmark by black dots in \cref{fig:summary}. Note that $\lambda_s$ is fixed to be $10^{-3}$ for illustration. In line with refs.~\cite{Caprini:2019egz, NANOGrav:2023hvm}, we use the broken power-law parametrization to depict the GW spectrum
\begin{eqnarray}
h^2_{} \Omega^{}_{\rm GW} = h^2_{} \Omega^{}_p \frac{(a+b)^c}{\left(b x^{-a / c}+a x^{b / c}\right)^c} \ ,
\label{eq:spec-par}
\end{eqnarray}
where $x := f/f_p$, and $a$, $b$ and $c$ are real and positive parameters. $a = 3$ is fixed in the low-frequency range due to causality, whereas $b$ and $c$ are chosen to be one based on the simulation results~\cite{Hiramatsu:2013qaa}. In \cref{fig:fit-DW2}, we exhibit our benchmark GW spectrum using the orange curve, while the gray violins denote the GW spectrum observed by the NANOGrav. One can find that the benchmark spectrum aligns with the NANOGrav 15-year result very well. With the help of eq.~\eqref{eq:spec-par}, we also estimate the testing capabilities of the future GW detectors ET~\cite{Punturo:2010zz}, LISA~\cite{LISA:2017pwj}, DECIGO~\cite{Kawamura:2020pcg}, $\mu$Ares~\cite{Sesana:2019vho}, SKA~\cite{Janssen:2014dka} and THEIA~\cite{Garcia-Bellido:2021zgu} by calculating the signal-to-noise ratio (SNR)~\cite{Maggiore:1999vm,Allen:1997ad}
\begin{equation}
    \varrho=\left[n_{\mathrm{det}} t_{\mathrm{obs}} \int_{f_{\min }}^{f_{\max }} d f\left(\frac{\Omega_{\text {signal }}(f)}{\Omega_{\text {noise }}(f)}\right)^2\right]^{1 / 2} \; ,
    \label{eq:SNR}
\end{equation}
where $n_{\rm det} = 1$ for auto-correlated detectors and  $n_{\rm det} = 2$ for cross-correlated detectors, $t_{\rm obs}$ represents the observational time, and $\Omega_{\text {noise }}$ denotes the noise spectrum expressed in terms of the GW energy density spectrum, which is essentially adopted from ref.~\cite{Schmitz:2020syl}. Integrating $(\Omega_{\rm signal}/\Omega_{\rm noise})^2$ over the sensitive frequency range of individual GW detectors, we obtain the SNRs for the GW spectra with different peak frequencies and energy densities. In \cref{fig:summary}, we mark the threshold SNR $\varrho = 10$ for different detectors by the solid-colored curves, which delineate the parameter ranges detectable by these GW detectors. In order to showcase the detection capabilities of these GW detectors, we further calculate the power-law-integrated sensitivity curves~\cite{Thrane:2013oya}, where we also choose $\varrho = 10$ as the threshold SNR. The results are also shown in \cref{fig:fit-DW2}. From \cref{fig:summary} and \cref{fig:fit-DW2}, we can find that our benchmark GW spectrum is within the detectable ranges of almost all the GW detectors except ET, which indicates that future GW detectors are very likely to validate the GW spectrum we predict by considering the DWs annihilation induced by the QG effects across a broader frequency range. It is also clear that these GW detectors will undoubtedly impose strong constraints on large $\Lambda_{\rm QG}$ in the future. In addition, the requirement that DWs should annihilate before BBN ($\tau_\text{DW}<t_{\rm ann}\lesssim 0.01~\text{s}$) also constrains the parameter space of $\Lambda_{\rm QG}$ and $v_s$, excluding the parameter range shaded by the gray color in \cref{fig:summary}.

\section{Conclusion}\label{sec:conclusion}

QG is of great interest to the broader community. As a continuation of our previous work~\cite{King:2023ayw}, in this paper, we discussed a simple model
inspired by string theory involving a fermionic DM candidate, together with another scalar responsible for creating DWs.
The two fields preserve discrete global $\Ztwo$ symmetries slightly broken by the non-perturbative effects of QG, which then leads to observable effects due to the DM decays and GWs from DWs annihilation. 

While the spirit of the approach is similar to the case of a scalar DM in \cite{King:2023ayw}, we encountered many new features specific to fermionic DM which makes the present study interesting. Indeed, the fermionic DM discussed in the present paper is a FIMP, which can be produced from out-of-equilibrium decays and achieve the correct relic density via the freeze-in mechanism. Also, we found that the mixing of the fermionic DM with SM neutrinos can have measurable effects in X/$\gamma$-ray, CMB and $0\nu\beta\beta$ decay observations, which can be used to constrain the QG scale. We found that there exist mutual constraints among the QG scale, the freeze-in scale and the reheating temperature. On the other hand, the predicted GW spectrum arising from the DWs annihilation may explain recent observations of the GW spectrum observed by several pulsar timing array projects and can be tested by future GW detectors. 

It has often been said that it is very difficult to explore QG effects in experiments. We have shown in this paper, however, that we can explore effective energy scales many orders above the Planck scale with current and future observations, and directly test theoretical ideas that are relevant at the highest energies, such as those originating from string theory.

\section*{Acknowledgements}

SFK and GW acknowledge the STFC Consolidated Grant ST/L000296/1 and SFK also acknowledge the European Union's Horizon 2020 Research and Innovation programme under the Marie Sklodowska-Curie grant agreement HIDDeN European ITN project (H2020-MSCA-ITN-2019//860881-HIDDeN). 
RR acknowledges financial support from the STFC Consolidated Grant ST/T000775/1. 
XW acknowledges the Royal Society as the funding source of the Newton International Fellowship.
MY was supported in part by the JSPS Grant-in-Aid for Scientific Research (No.\ 19H00689, 19K03820, 20H05860, 23H01168), and by JST, Japan (PRESTO Grant No. JPMJPR225A, Moonshot R\&D Grant No.\ JPMJMS2061).

\bibliographystyle{apsrev4-1}
\bibliography{refs}
\end{document}